\documentstyle[graphicx, 12pt]{article}
\begin{document}

\small
\hoffset=-1truecm
\voffset=-2truecm
\title{\bf The deflection angle of a gravitational source with global monopole in the strong field limit}
\author{Hongbo Cheng\footnote {E-mail address: hbcheng@sh163.net} \hspace {1cm} Jingyun Man\\
Department of Physics, East China University of Science and
Technology,\\ Shanghai 200237, China\\
The Shanghai Key Laboratory of Astrophysics, Shanghai 200234,
China}

\date{}
\maketitle

\begin{abstract}
We investigate the gravitational lensing effect in the strong
field background around the Schwarzschild black hole with
extremely small mass and solid deficit angle subject to the global
monopole by means of the strong field limit issue. We obtain the
angular position and magnification of the relativistic images and
show that they relate to the global monopole parameter $\eta$. We
discuss that with the increase of parameter $\eta$, the minimum
impact parameter $u_{m}$ and angular separation $s$ increase and
the relative magnification $r$ decreases. We also find that $s$
grows extremely greatly as the increasing parameter $\eta$ becomes
large enough. The deflection angle will become larger when the
parameter $\eta$ grows. The effect from the solid deficit angle is
the dependence of angular position, angular separation, relative
magnification and deflection angle on parameter $\eta$, which may
offer a way to characterize some possible distinct signatures of
the Schwarzschild black hole with a solid deficit angle associated
with global monopole.
\end{abstract}
\vspace{6cm} \hspace{1cm} PACS number(s): 95.30.Sf, 04.70.-s

\newpage

\noindent \textbf{I.\hspace{0.4cm}Introduction}

The deflection of electromagnetic radiation in a gravitational
field leads the phenomena referred as gravitational lensing, while
a massive object which brings about the detectable deflection is
called a gravitational lens. Certainly the gravitational lensing
is an important application of general relativity [1]. The basic
theory of gravitational lensing has been developed greatly and
applied widely [2-7]. As a useful tool in astrophysics, the
gravitational lensing helps us to investigate the distant stars no
matter whether they are bright or dim and to verify the gravity
theories.

The various topological defects including domain walls, cosmic
strings and monopoles subject to the topology of the vacuum
manifold produced during the vacuum phase transition in the early
Universe [8, 9]. A global monopole is a spherical symmetric
gravitational topological defect formed within the phase
transition of a system consisting of a self-coupling scalar field
triplet whose original global $O(3)$ symmetry is spontaneously
broken to $U(1)$ and its gravitational effects lead to the
peculiarity like deficit solid angle, which makes all light rays
to be deflected by the same angle [10]. When a global monopole
lives between an observer and a source, the observer will see two
point images separated by an angle related to the deficit angle.
More efforts have been contributed to the topics related to the
global monopoles. It was shown that the global monopoles provide
their relevant contributions to the vacuum polarizations in
several kinds of backgrounds [11, 12]. The global monopoles were
discussed in the spacetimes with cosmological constant and the
relations between the models and spacetime topology were obtained
[13, 14]. The models of composite monopoles have been also
considered and the interactions between them were shown [15].

The gravitational lensing is a useful technique to probe the
galaxies and the Universe [5, 6]. The technique can also help us
to study the dark matter [16, 17] and dark energy [18]. In general
the deflection angle of light passing close to a compact and
massive source is expressed in integral forms, so it is difficult
to discuss the detailed relations between the angle and the
gravitational source. Alternatively we perform the calculation of
the integral expressions in the limiting cases such as weak field
approximation or strong ones. Most of approaches of gravitational
lensing have been developed in the weak field approximation with
the first non-null term in the expansion of the deflection angle
[5, 6]. When the light passes very close to a heavy compact body
such as a black hole, the amount of the deflection of the light
must be quite large because it depends on the mass. In this region
the gravitational field is strong and an infinite series of images
produced very close to the black hole can be observed. Here the
weak field approximation is not valid any more, so the strong
gravitational lensing has attracted wide attention. For example,
the strong gravitational lensing was treated in a Schwarzschild
black hole [19, 20], gravitational source with naked singularities
[21], a Reissner-Nordstrom black hole [22], a GMGHS charged black
hole [23], a spining black hole [24, 25], a braneworld black hole
[26, 27], an Einstein-Born-Infeld black hole [28], a black hole in
Brans-Dicke theory [29], a Barriola-Vilenkin monopole [30] and the
deformed Horava-Lifshitz black hole [31], etc.. According to
COSMOS survey there could be strong-lensing candidates [32, 33].
As mentioned previously, the calculation for integral expression
is burden, the method to deal with the derivative in the strong
gravitational field cases is necessary. It should be pointed out
that Bozza put forward an analytical method which can help us to
distinguish the nature of environment around various types of
black holes. This method showed that there exists a logarithmic
divergence of the deflection angle in proximity of the photon
sphere [34, 35] and has been improved [36-39].

In this paper we shall explore the gravitational lensing on the
Schwarzschild black hole swallowing a global monopole or
equivalently the gravitational source with deficit solid angle.
Although the gravitational lens equation for the Barriola-Vilenkin
monopole was considered [30], it is necessary to study the
influence from the monopole model parameters related to the
deficit solid angle on the deflection angle in the case of massive
object further and in detail. We wonder whether the deficit solid
angle generates the other distinct features besides it inspires
two images mentioned in Ref. [10]. We plan to research on the
deflection angle for light rays propagating in the background
around the massive source with Barriola-Vilenkin monopole by means
of Bozza's device. We plot the dependence of the observational
gravitational lensing parameters on the deficit solid angle from
the global monopole. We compare the properties of gravitational
lensing in the case of massive source containing Barriola-Vilenkin
monopole with those of Schwarzschild metrics. At last we list our
results.

\vspace{0.8cm} \noindent \textbf{II.\hspace{0.4cm}The deflection
angle of a massive source with global monopole}

The Lagrangian density describing the simplest model of a global
monopole is [9, 10],

\begin{equation}
{\cal
L}=\frac{1}{2}(\partial_{\mu}\phi^{a})(\partial^{\mu}\phi^{a})
-\frac{1}{4}\lambda(\phi^{a}\phi^{a}-\eta^{2})^{2}
\end{equation}

\noindent Here $\lambda$ and $\eta$ are parameters in this model.
Introducing the ansatz of a self-coupling scalar field triplet
like $\phi^{a}=f(r)\frac{x^{a}}{r}$ and coupling this matter field
with the Einstein equations, a static and spherically symmetric
black hole solution was given by the line element,

\begin{equation}
ds^{2}=A(r)dt^{2}-B(r)dr^{2}-r^{2}(d\theta^{2}+\sin^{2}\theta
d\varphi^{2})
\end{equation}

\noindent where

\begin{equation}
A=B^{-1}=1-8\pi G\eta^{2}-\frac{2GM}{r}
\end{equation}

\noindent describing the background far from the monopole's core.
$G$ is the Newton constant. The mass parameter is $M\sim M_{core}$
and $M_{core}$ is very small. The metric of this black hole has
the event horizon like,

\begin{equation}
r_{h}=\frac{2GM}{1-8\pi G\eta^{2}}
\end{equation}

\noindent The parameter $\eta$ in functions $A(r)$ and $B(r)$ will
make a great deal of influence on gravitational lensing in the
strong field limit.

According to the Ref. [5, 6, 34], we choose that both the observer
and the gravitational source lie in the equatorial plane with
conditions $\theta=\frac{\pi}{2}$. The whole trajectory of the
photon is subject to the same plane. On the equatorial plane the
metric (2) becomes,

\begin{equation}
ds^{2}=A(r)dt^{2}-B(r)dr^{2}-C(r)d\varphi^{2}
\end{equation}

\noindent where

\begin{equation}
C(r)=r^{2}
\end{equation}

\noindent The deflection angle for the electromagnetic ray coming
from remote place can be expressed as a function of the closest
approach [40],

\begin{equation}
\alpha(r_{0})=I(r_{0})-\pi
\end{equation}

\noindent and

\begin{equation}
I(r_{0})=2\int_{r_{0}}^{\infty}\frac{\sqrt{B(r)}}{\sqrt{C(r)}\sqrt{\frac{C(r)A(r_{0})}{C(r_{0})A(r)}}-1}dr
\end{equation}

\noindent where $r_{0}$ is the minimum distance from the photon
trajectory to the source. That the denominator of expression (8)
is equal to the zero leads to deflection angle to be divergent,
which means that the photon is captured by the source instead of
moving towards the observer. We call the largest root of equation
$\frac{C'(r)}{C(r)}=\frac{A'(r)}{A(r)}$ the radius of the photon
sphere [21, 41]. The closest approach distant $r_{0}$ must be
larger than the radius of photon sphere or the photon can not
escape from the gravitational source. The radius of the photon
sphere in the global monopole metric can be given by,

\begin{equation}
r_{m}=\frac{3GM}{1-8\pi G\eta^{2}}
\end{equation}

\noindent and is larger than the event horizon (4). The photon
sphere radius (9) will recover to that of Schwarzschild spacetime
as the parameter $\eta=0$. We wonder how the parameter $\eta$ of
global monopole generate the effects on gravitational lensing. In
Ref. [34] Bozza put forward an important expression of deflection
angle for the metric which is assumed asymptotically flat. Having
followed the procedure of Ref. [34], in the strong field limit we
re-derive the deflection angle of the metric (5) whose element
satisfies $\lim_{r\longrightarrow\infty}A(r)\neq 1$. Here let
$l_{0}=\lim_{r\longrightarrow\infty}A(r)$ and follow the procedure
of Ref. [34]. We can generalize Bozza's work to denote the
deflection angle in the strong field limit as follow [34],

\begin{equation}
\alpha(\theta)=-\bar{a}\ln(\frac{\theta
D_{OL}}{u_{m}}-1)+\bar{b}+O(u-u_{m})
\end{equation}

\noindent In order to apply the expression to the case with global
monopole, we define some functions as follow,

\begin{equation}
R(z,
r_{0})=\frac{2\sqrt{B(r)A(r)}}{C(r)A'(r)}(l_{0}-A(r_{0}))\sqrt{C(r_{0})}
\end{equation}

\begin{equation}
f(z,
r_{0})=\frac{1}{\sqrt{A(r_{0})-[(l_{0}-A(r_{0}))z+A(r_{0})]\frac{C(r_{0})}{C(r)}}}
\end{equation}

\begin{equation}
f_{0}(z, r_{0})=\frac{1}{\sqrt{\alpha z+\beta z^{2}}}
\end{equation}

\begin{equation}
I_{D}(r_{0})=\int_{0}^{1}R(0, r_{m})f_{0}(z, r_{0})dz
\end{equation}

\begin{equation}
I_{R}(r_{0})=\int_{0}^{1}[R(z, r_{0})f(z, r_{0})-R(0,
r_{m})f_{0}(z, r_{0})]dz
\end{equation}

\noindent where

\begin{equation}
z=\frac{A(r)-A(r_{0})}{l_{0}-A(r_{0})}
\end{equation}

\begin{equation}
\alpha=\frac{l_{0}-A(r_{0})}{C(r_{0})A'(r_{0})}(C'(r_{0})A(r_{0})-C(r_{0})A'(r_{0}))
\end{equation}

\begin{equation}
\beta=\frac{(l_{0}-A(r_{0}))^{2}}{2C^{2}(r_{0})A'^{3}(r_{0})}[2C_{0}C'_{0}A'^{2}_{0}
+(C_{0}C''_{0}-2C'^{2}_{0})A_{0}A'_{0}-C_{0}C'_{0}A_{0}A''_{0}]
\end{equation}

\noindent with $A_{0}=A(r_{0})$, $C_{0}=C(r_{0})$. It should be
pointed out that our results will recover to be those of Ref. [34]
by Bozza if $l_{0}=1$. In Eq. (10) $D_{OL}$ means the distance
between observer and gravitational lens. By conservation of the
angular momentum around the source, the closest approach distance
is related to the impact parameter $u$ defined as
$u=\sqrt{\frac{C(r_{0})}{A(r_{0})}}$ [40]. Further the minimum
impact parameter corresponding to $r_{0}=r_{m}$ is certainly
$u_{m}=u\mid_{r_{0}=r_{m}}$. The strong field limit coefficients
$\bar{a}$ and $\bar{b}$ can be expressed as,

\begin{equation}
\bar{a}=\frac{R(0, r_{m})}{2\sqrt{\beta_{m}}}
\end{equation}

\begin{equation}
\bar{b}=-\pi+I_{R}(r_{m})+\bar{a}\ln\frac{2\beta_{m}}{A(r_{m})}
\end{equation}

\noindent where

\begin{equation}
\beta_{m}=\beta|_{r_{0}=r_{m}}
\end{equation}

\noindent According to Barriola-Vilenkin-monopole type metric (3),
here we find $l_{0}=1-8\pi G\eta^{2}$. According to the
discussions above we apply the device to the case of massive
source involving global monopole with metric (2) to obtain that

\begin{equation}
R(z, r_{0})=2
\end{equation}

\begin{equation}
f(z, r_{0})=[-\frac{2GM}{r_{0}}z^{3}-(1-8\pi
G\eta^{2}-\frac{6GM}{r_{0}})z^{2}+2(1-8\pi
G\eta^{2}-\frac{3GM}{r_{0}})z]^{-\frac{1}{2}}
\end{equation}

\begin{equation}
f_{0}(z, r_{0})=[-(1-8\pi
G\eta^{2}-\frac{6GM}{r_{0}})z^{2}+2(1-8\pi
G\eta^{2}-\frac{3GM}{r_{0}})z]^{-\frac{1}{2}}
\end{equation}

\noindent It is clear that the function $R(z, r_{0})$ is regular
and the function $f(z, r_{0})$ diverges as $z$ vanishes. According
to the definition above and Eq. (19) and (20), we derive the
coefficients $\bar{a}$, $\bar{b}$ and minimum impact parameter
$u_{m}$ of the deflection angle,

\begin{equation}
\bar{a}=\frac{1}{\sqrt{1-8\pi G\eta^{2}}}
\end{equation}

\begin{equation}
\bar{b}=\frac{4}{\sqrt{1-8\pi
G\eta^{2}}}\ln(3-\sqrt{3})+\frac{1}{\sqrt{1-8\pi
G\eta^{2}}}\ln6-\pi
\end{equation}

\begin{equation}
u_{m}=(\frac{3}{1-8\pi G\eta^{2}})^{\frac{3}{2}}GM
\end{equation}

\noindent In the strong field limit the deflection angle of
massive source with deficit angle is,

\begin{eqnarray}
\alpha(\theta)=-\frac{1}{\sqrt{1-8\pi G\eta^{2}}}\ln(\frac{\theta
D_{OL}}{(\frac{3}{1-8\pi
G\eta^{2}})^{\frac{3}{2}}GM}-1)+\frac{4}{\sqrt{1-8\pi
G\eta^{2}}}\ln(3-\sqrt{3})\nonumber\\+\frac{1}{\sqrt{1-8\pi
G\eta^{2}}}\ln6-\pi\hspace{6cm}
\end{eqnarray}

\noindent All of the coefficients $\bar{a}$, $\bar{b}$ and $u_{m}$
are larger than those of the standard Schwarzschild black hole
without global monopole. The expressions (25)-(27) also show that
these coefficients become larger as the global monopole parameter
$\eta$ increases. This enables us to distinguish a massive source
with global monopole from a Schwarzschild black hole in virtue of
strong field gravitational lensing even if the two gravitational
source have the same mass.

In order to compare our results with the observable evidence we
relate the position and the magnification to the strong field
limit coefficients. The lens equation in the strong field limit
reads [39],

\begin{equation}
\beta=\theta-\frac{D_{LS}}{D_{OL}}\bigtriangleup\alpha_{n}
\end{equation}

\noindent where $D_{LS}$ represents the distance between the lens
and the source. Here $D_{OS}=D_{OL}+D_{LS}$. $\beta$ denotes the
angular separation between the source and the lens. $\theta$ is
the angular separation between the lens and the image. Once we
subtract all the loops made by the photon, the offset of the
deflection angle is expressed as
$\bigtriangleup\alpha_{n}=\alpha(\theta)-2n\pi$. Because of
$u_{m}\ll D_{OL}$ the position of the $n$-th relativistic image
can be approximated as,

\begin{equation}
\theta_{n}=\theta_{n}^{0}+\frac{u_{m}e_{n}(\beta-\theta_{n}^{0})D_{OS}}{\bar{a}D_{LS}D_{OL}}
\end{equation}

\noindent where

\begin{equation}
e_{n}=e^{\frac{\bar{b}-2n\pi}{\bar{a}}}
\end{equation}

\noindent while the term in the right-hand side of Eq. (30) is
much smaller than $\theta_{n}^{0}$ and we introduce
$\theta_{n}^{0}$ as $\alpha(\theta_{n}^{0})=2n\pi$. The
magnification of $n$-th relativistic image is the inverse of the
Jacobian evaluated at the position of the image and is obtained
as,

\begin{eqnarray}
\mu_{n}=\frac{1}{(\frac{\beta}{\theta})\frac{\partial\beta}{\partial\theta}}|_{\theta_{n}^{0}}
\nonumber\hspace{1cm}\\=\frac{u_{m}e_{n}(1+e_{n})D_{OS}}{\bar{a}\beta
D_{LS}D_{OL}^{2}}
\end{eqnarray}

\noindent In the limit $n\longrightarrow\infty$ the asymptotic
position of approached by a set of images $\theta_{\infty}$
relates to the minimum impact parameter as,

\begin{equation}
u_{m}=D_{OL}\theta_{\infty}
\end{equation}

\noindent As an observable the angular separation between the
first image and the others is defined as,

\begin{equation}
s=\theta_{1}-\theta_{\infty}
\end{equation}

\noindent where $\theta_{1}$ represents the outermost image in the
situation that the outermost one is thought as a single image and
all the remaining ones are packed together at $\theta_{\infty}$.
As another observable the ratio of the flux from the first image
and the flux of all the other images is,

\begin{equation}
r=\frac{\mu_{1}}{\sum_{n=2}^{\infty}\mu_{n}}
\end{equation}

\noindent According to $e^{\frac{2\pi}{\bar{a}}}\gg1$ and
$e^{\frac{\bar{b}}{\bar{a}}}\sim1$, the above formulae (34) and
(35) can be simplified as

\begin{equation}
s=\theta_{\infty}e^{\frac{\bar{b}-2\pi}{\bar{a}}}
\end{equation}

\begin{equation}
r=e^{\frac{2\pi}{\bar{a}}}
\end{equation}

\noindent It is significant that the strong field limit
coefficients such as $\bar{a}$, $\bar{b}$ and $u_{m}$ are directly
connected to the observables like $r$ and $s$. It is then possible
for us to discern a massive object with global monopole from the
standard Schwarzschild black hole by means of strong field
gravitational lensing.

\vspace{0.8cm} \noindent \textbf{III.\hspace{0.4cm}Numerical
estimation of observables in the strong field limit}

It is necessary to estimate the numerical values of the
coefficients and observables of gravitational lensing in the
strong field limit to explore the influence from global monopole
existing in the gravitational source. Our results will recover to
those of standard Schwarzschild black hole spacetime as $\eta=0$
although the mass is tiny. We calculate these coefficients and
observables according to equations above and show their dependence
on the model parameter $\eta$ relating to the global monopole in
the figures. From Fig. 1 we discover that the increasing model
parameter leads the minimum impact parameter $u_{m}$ greater
although the source mass may be extremely tiny. Fig. 2 exhibits
that the two strong gravitational lensing coefficients $\bar{a}$
and $\bar{b}$ both become larger with the increase of model
parameter $\eta$. In the case of pure Schwarzschild black hole the
two coefficients keep constant. These can help us to find whether
the metric of the gravitational source has solid deficit angle
subject to the global monopole. Fig. 3 and Fig. 4 also show that
with the increase of parameter $\eta$, the angular separation $s$
becomes larger while the relative magnification $r$ decreases in
the case of keeping the total mass constant. It should be pointed
out that the angular separation $s$ increases extremely quickly as
the parameter $\eta$ becomes sufficiently large. In Fig. 5 we
demonstrate the relation between the deflection angle and the
parameter $\eta$ for various values of $\frac{\theta D_{OL}}{GM}$
according to Eq. (28) and the curves are similar. It is evident
that the deflection angle becomes much greater when the increasing
global monopole parameter $\eta$ is sufficiently large within the
region $G\eta^{2}\in(0, \frac{1}{8\pi})$ in Fig. 5. The
model-dependent deflection angle is certainly larger than that of
the standard Schwarzschild black hole. In all of figures mentioned
above the deviations subject to the global monopole are evident,
so the strong gravitational lensing is efficient to explore the
monopole. The influence from the global monopole is greater, which
helps us to identify the nature of the massive source that has the
deficit solid angle even the mass is completely tiny.

\vspace{0.8cm} \noindent \textbf{IV.\hspace{0.4cm}Summary}

In this paper we study the gravitational lensing in the strong
field limit in the Schwarzschild black hole spacetime with solid
deficit angle owing to global monopole. The global monopoles
contained by a massive source bring about strong influence on the
observation. It is found that the global monopole leads the photon
sphere radius of gravitational source to be larger. We indicate
that in the case of negligible mass the increase of global
monopole model parameter $\eta$ leads the minimum impact parameter
$u_{m}$ to increase and the relative flux $r$ to decrease. We find
that the angular separation $s$ becomes larger as the parameter
$\eta$ increases and increases extremely quickly as the increasing
parameter $\eta$ goes to large enough. It should be pointed out
that the deflection angle grows with the increase in parameter
$\eta$. The sufficiently large $\eta$ will lead much more manifest
deflection of light. It is interesting that the deficit solid
angle shows some other distinct characteristics besides two images
presented in Ref. [10]. The effect from solid deficit angle on the
astronomical observations is manifest, which provides us a way to
probe the spacetime property such as solid deficit angle resulting
from global monopole. The gravitational lensing in the strong
field limit is a potentially powerful tool to research on the
nature of various black holes.

\vspace{1cm}
\noindent \textbf{Acknowledge}

This work is supported by NSFC No. 10875043 and is partly
supported by the Shanghai Research Foundation No. 07dz22020.

\newpage
\begin{figure}
\setlength{\belowcaptionskip}{10pt} \centering
  \includegraphics[width=15cm]{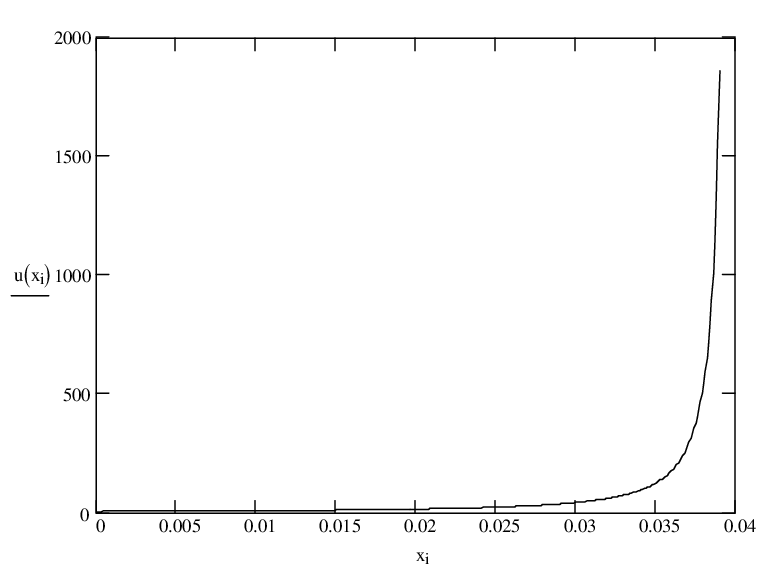}
  \caption{The dependence of the minimum impact parameter $u_{m}$ in unit of $GM$ for strong field limit in the Schwarzschild
  black hole with global monopole on model parameter $x=G\eta^{2}$.}
\end{figure}

\newpage
\begin{figure}
\setlength{\belowcaptionskip}{10pt} \centering
  \includegraphics[width=15cm]{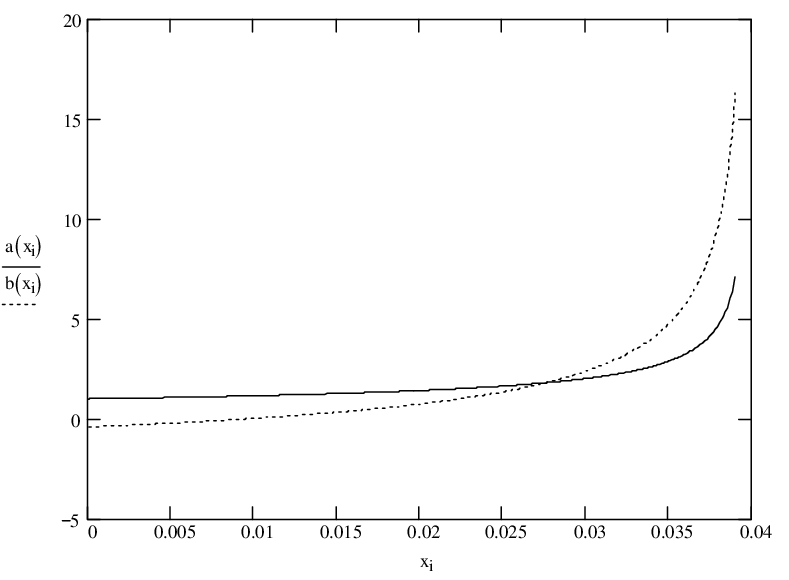}
  \caption{The solid and dot curves corresponds to the dependence of the coefficient of
  strong field limit $\bar{a}$ and $\bar{b}$ respectively in the Schwarzschild black hole
  with global monopole on model parameter $x=G\eta^{2}$.}
\end{figure}

\newpage
\begin{figure}
\setlength{\belowcaptionskip}{10pt} \centering
  \includegraphics[width=15cm]{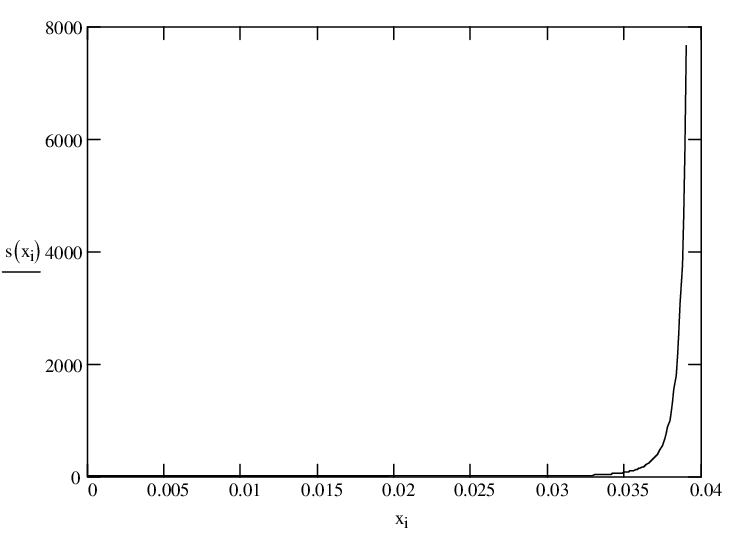}
  \caption{The dependence of the angular separation $s$ in the Schwarzschild
  black hole with global monopole on model parameter $x=G\eta^{2}$ in unit of $\theta_{\infty}$,
  the asymptotic position of approached by a set of images.}
\end{figure}

\newpage
\begin{figure}
\setlength{\belowcaptionskip}{10pt} \centering
  \includegraphics[width=15cm]{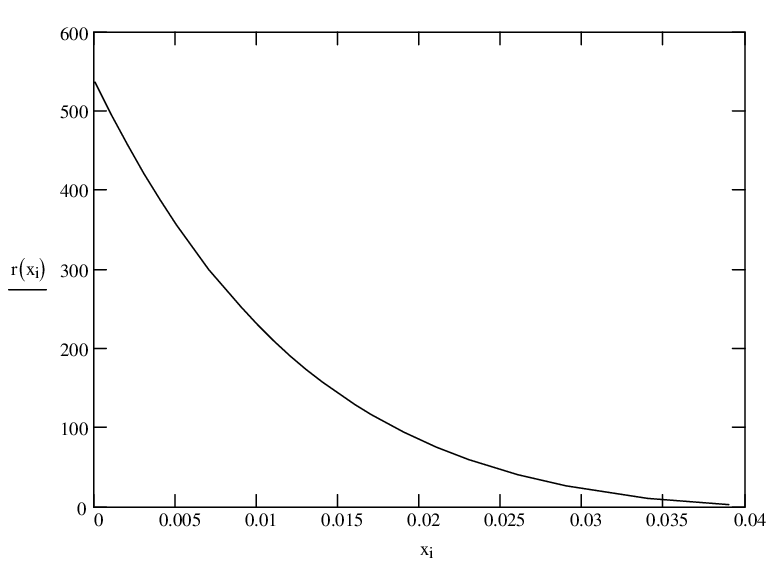}
  \caption{The dependence of the relative flux $r$ in the Schwarzschild
  black hole with global monopole on model parameter $x=G\eta^{2}$.}
\end{figure}

\newpage
\begin{figure}
\setlength{\belowcaptionskip}{10pt} \centering
  \includegraphics[width=15cm]{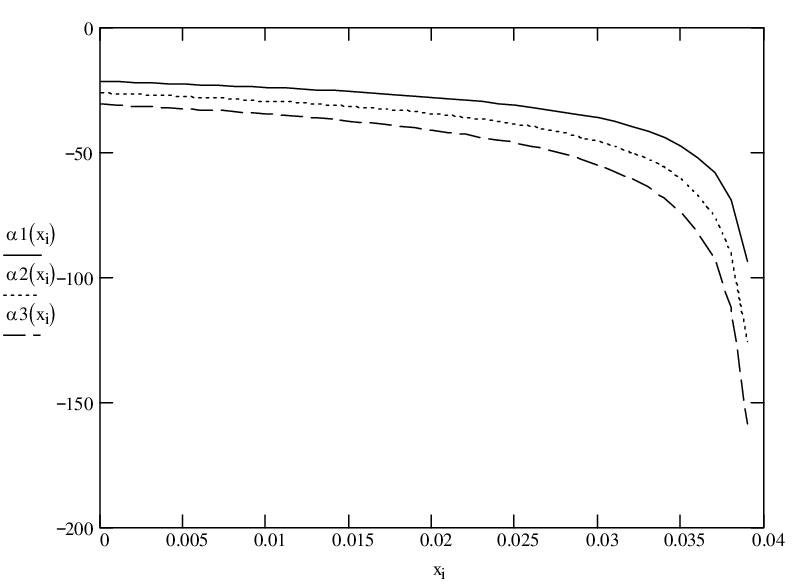}
  \caption{The solid, dot and dashed curves correspond to the dependence of the deflection angle
  $\alpha$ in the Schwarzschild
  black hole with global monopole on model parameter $x=G\eta^{2}$
  for $\frac{\theta D_{OL}}{GM}=10^{10}, 10^{12}, 10^{14}$ respectively.}
\end{figure}

\end{document}